\documentclass[aps,prb,preprint,superscriptaddress,longbibliography]{revtex4-2}
\usepackage{geometry}
\usepackage{amsmath,amssymb}
\usepackage{siunitx}
\usepackage{graphicx}
\usepackage{bm}
\usepackage[T1]{fontenc}
\usepackage{textcomp,gensymb}
\usepackage[colorlinks=true,linkcolor=blue,citecolor=blue,urlcolor=blue]{hyperref}
\usepackage[caption=false]{subfig} 

\begin{document}

\title{On the Practicability of Ceramic-Tiled Walls for Sound Absorption by Tuning Cavities}

\author{Ozgur T. Tugut${}^{\dagger}$}
\affiliation{Université Marie et Louis Pasteur, Institut FEMTO-ST, CNRS, 25000 Besançon, France}

\author{Brahim Lemkalli${}^{\dagger, \;\ast}$}
\affiliation{Université Marie et Louis Pasteur, Institut FEMTO-ST, CNRS, 25000 Besançon, France}

\author{Qingxiang Ji}
\affiliation{Université Marie et Louis Pasteur, Institut FEMTO-ST, CNRS, 25000 Besançon, France}

\author{Mahmoud Addouche}
\affiliation{Université Marie et Louis Pasteur, Institut FEMTO-ST, CNRS, 25000 Besançon, France}

\author{Benjamin Vial}
\affiliation{Mathematics Department, Imperial College London, SW7~2AZ, London, UK}

\author{Sébastien Guenneau}
\affiliation{UMI 2004 Abraham de Moivre-CNRS, Imperial College London, SW7~2AZ, London, UK}
\affiliation{The Blackett Laboratory, Physics Department, Imperial College London, SW7~2AZ, London, UK}

\author{Richard Craster}
\affiliation{Mathematics Department, Imperial College London, SW7~2AZ, London, UK}
\affiliation{UMI 2004 Abraham de Moivre-CNRS, Imperial College London, SW7~2AZ, London, UK}

\author{Claudio Bizzaglia}
\affiliation{IRIS, Via Guido Reni 2E, 42014 Castellarano, Italy}

\author{Bogdan Ungureanu}
\affiliation{Mathematics Department, Imperial College London, SW7~2AZ, London, UK}
\affiliation{The Blackett Laboratory, Physics Department, Imperial College London, SW7~2AZ, London, UK}
\affiliation{METAMAT Limited, 937 Chelsea C, Sloane Avenue, SW3~3EU, London, UK}

\author{Muamer Kadic}
\affiliation{Université Marie et Louis Pasteur, Institut FEMTO-ST, CNRS, 25000 Besançon, France}

\date{\today}

\begin{abstract}
we present the practicality of structuring ceramic tiles for enhancing sound absorption on rigid walls.
%We apply a methodology allowing isospectral tuning of reference cavities with either smaller or larger ones, enforcing their spectra to coincide.
The cornerstone of our methodology is to structure walls with cavities so that walls effectively behave as heterogeneous absorbing surfaces over a large frequency bandwidth.
%\cite{lenz2023transformation, cominelli2024isospectral}. 
Using this approach, ceramic tiled walls are developed by integrating tuned cavity structures based on Helmholtz resonators. Such a design leverages the empty joints between tiles to form resonator necks, while the space between the ceramic tiles and the wall acts as the resonator chambers. By arranging these resonators in a spatially graded array, we achieve broadband sound absorption which targets low-frequency noise generated by impacts, footsteps and ambient sources. This makes the system highly suitable for practical architectural applications. The study encompasses the entire process, from numerical modeling and analytical formulation to the fabrication and mounting of resonant tiles, followed by experimental validation, clearly demonstrating the effectiveness of the proposed solution in real-world conditions. The findings highlight the strong potential of this approach for practical tiled room acoustic treatment and noise mitigation.
\end{abstract}

\maketitle
\vspace{-1em}
\noindent${}^{\dagger}$~These authors contributed equally to this work.\\
\noindent${}^{\ast}$~Corresponding author: \href{mailto:lemkallibrahim@gmail.com}{lemkallibrahim@gmail.com}
% \linenumbers
\section{Introduction}
Sound absorption is a crucial aspect of acoustic design in architectural and industrial applications, aimed at controlling noise and improving auditory environments. Conventional sound absorption methods primarily rely on porous materials \cite{fengshan2020equivalent, dong2024recent} and micro-perforated structures \cite{ning2016acoustic, laly2019sensitivity}, both of which typically require a structural thickness on the order of the acoustic wavelength. To effectively absorb low-frequency sound, which has longer wavelengths, these materials must be significantly thicker, thereby limiting their practical applications \cite{ren2022compact}. In addition, these absorbers usually result in an imperfect impedance match with the incident wave \cite{li2016sound}. 

Traditional building materials typically rely on their intrinsic material composition, such as density and porosity, to enhance global acoustic properties. In contrast, the performance of metamaterials is primarily governed by their geometric design, structural configuration, or stiffness distribution, allowing them to be tailored into a wide range of functional bulk materials \cite{wang2022nonlocal,craster2023mechanical,huang2024sound}. To address the limitations of conventional materials, acoustic metamaterials have emerged as a promising solution for building insulation, gaining significant attention due to their exceptional wave-control capabilities in complex acoustic environments \cite{craster2023mechanical}. These metamaterials are artificially engineered to exhibit unique properties not commonly found in natural materials, spanning multiple disciplines including electromagnetics, optics, solid-state physics, and acoustics. Acoustic metamaterials are typically composed of periodic or aperiodic substructures that enable targeted or unconventional acoustic behavior, often leveraging local resonance phenomena by leveraging subwavelength resonances, wave interference effects or non-localities \cite{wang2022nonlocal,iglesias2021three,gao2024design}. 

Recent advances in the field of acoustic metamaterials \cite{wang2022nonlocal,ji2021designing} have paved the way for innovative solutions to achieve extraordinary effects such as acoustic cloaking \cite{kadic2015experiments,craster2023mechanical}, wave focusing and negative refraction \cite{cummer2007one,chen2007acoustic,chong2025design,chen2025broadband,kaina2015negative}. Beyond these new capabilities, acoustic metamaterials present promising alternatives to conventional sound-absorbing materials by enabling effective sound absorption at sub-wavelength dimensions. Various acoustic metamaterial designs have been proposed, such as membrane-type absorbers \cite{dong2025novel,li2023sound}, Helmholtz cavities \cite{garza2024metasurfaces,lemkalli2023bi, lemkalli2023lightweight,jimenez2016ultra,almeida2021low}, spatial coil resonators\cite{sun2024sound,yuan2022tunable}, neck-integrated Helmholtz cavities \cite{huang2024tunable,lemkalli2023innovative} and hierarchical porous structures\cite{zhou2025broadband,chen2023functionally}. The search for acoustic metamaterials combining compactness and broadband low-frequency absorption remains a crucial and ongoing area of research.
\begin{figure*}[h]
    \centering
    \includegraphics[width=0.6\linewidth]{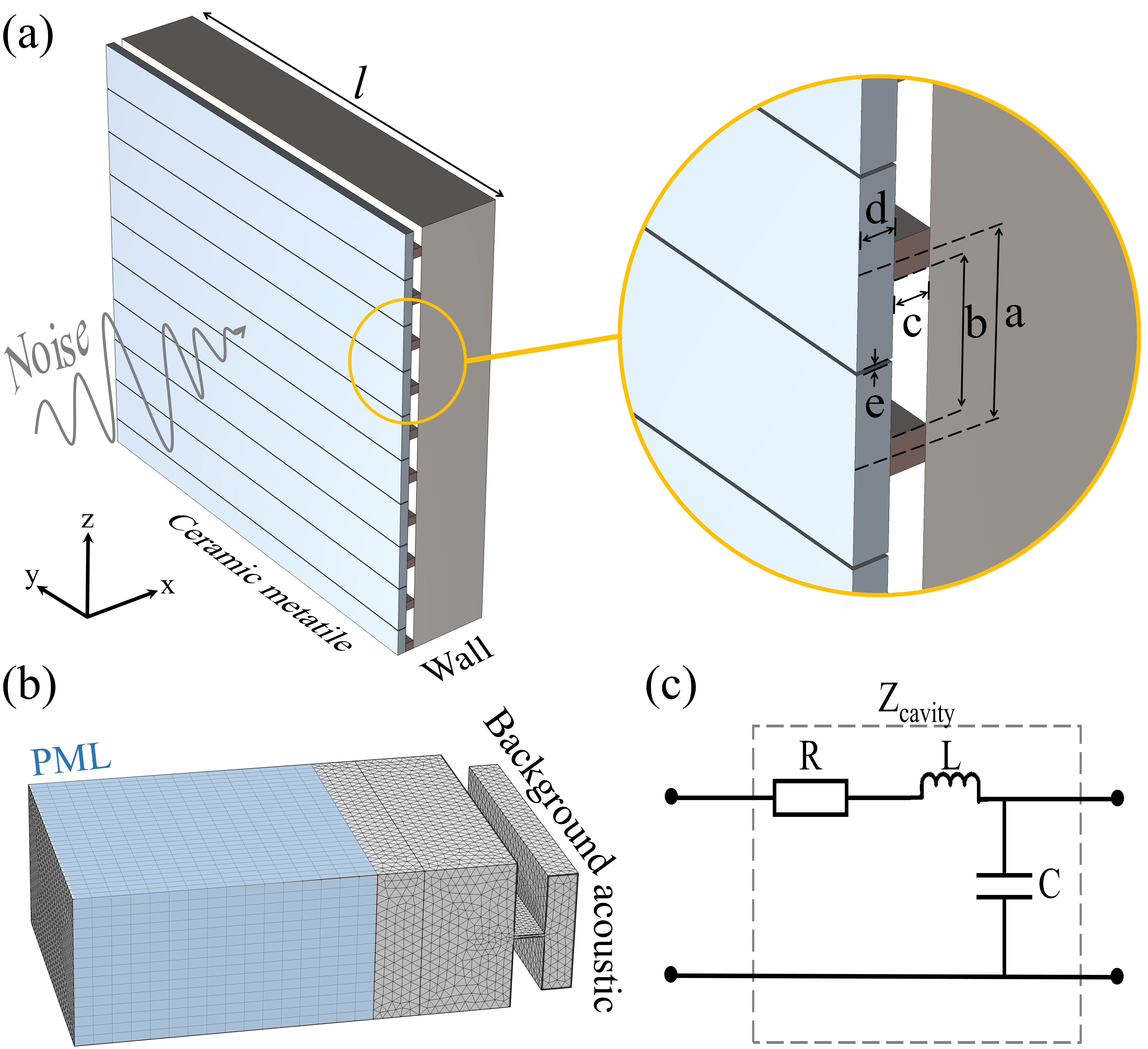}
    \caption{The ceramic metatile wall, which is composed of tuning cavity structures integrated into ceramic subtiles fixed to the wall using wooden beams. (a) Overview of the design (right) of the structural design of the metatile wall along with a zoom-in view of the unit cell (left). With, b is the height of the tuning cavity, c is the depth of the cavity, and e is the width of the cavity slits. The ceramic tiles have a thickness $d=1$ \si{cm} and a width of $5$ \si{cm}, while the wood beams measure $1$ \si{cm} in both width and depth. The structure is periodic in the $z$-direction, with a periodicity constant denoted by $a$ and the incident wave propagates along the $x$-direction. (b) Schematic view of the numerical model {with Perfectly Matched Layers (PML) in blue color and elastic medium in grey color.} (c) Equivalent {resistor, inductor, and capacitor (RLC)} circuit model of the unit cell {(an impedance cavity $Z_{\rm cavity}$) }.}
    \label{Figure1}
\end{figure*}

On the other hand, ceramic tiles are a widely used construction material applied in various parts of buildings, including household roofs, walls, floors, and partitions \cite{berto2007ceramic}. In modern architecture, wall tiles are typically designed for durability, abrasion resistance, water impermeability, ease of cleaning, and aesthetic appeal \cite{sudol2021makes}. However, in terms of acoustic absorption performance, ceramic tiles are not effective at reducing noise or sound absorption. Due to their high density and rigid surface, they exhibit poor sound absorption and high sound reflectivity, making them less suitable for applications requiring acoustic insulation in walls.

To address these challenges, we investigate tuning cavity structures \cite{lenz2023transformation,cominelli2024isospectral} as an effective strategy for achieving broadband sound absorption while maintaining the compactness, durability, ease of cleaning, and weather resistance of ceramic tiled walls. The core concept involves incorporating Helmholtz resonators, which are well-known for their ability to absorb low-frequency sound, directly into ceramic subtile assemblies. The gaps between adjacent periodic subtiles function as resonator necks or slits, while the air cavities formed between the subtiles and the wall act as the primary tuning cavities. By arranging an inhomogeneous array of such tuning cavities, broadband sound absorption can be achieved, allowing for efficient mitigation of audible noise over a wide low-frequency range. This approach offers a practical and architecturally seamless solution for noise control, utilizing the inherent structural advantages of ceramic tiles while integrating advanced acoustic functionality.

Therefore, in this study, we present a comprehensive investigation of tuning cavity structures, including numerical modeling, analytical approaches, and experimental validation. We demonstrate their effectiveness in real-world building applications, showing that this design enables high-performance sound absorption without compromising the aesthetic or structural integrity of ceramic wall systems.

\section{Design of the tuning cavity structure} 

The design method of the tuning cavity structures and the geometric parameters of their constituent units are introduced in this section. In addition, the methods for sound absorption calculations, both numerical and analytical, as well as the fabrication of the experimental sample and the measurement setup, are also discussed.
\subsection{Materials and design}
The structure of the metatile wall, based on tuned cavity units, is illustrated in \autoref{Figure1}(a). It consists of a series of ceramic subtiles of width $a$, separated by slits of width $e$. A gap of width $c$ between the ceramic subtiles and the wall forms a cavity that functions similarly to a Helmholtz resonator with the material properties shown in Table \ref{tab1}. This cavity plays a crucial role in absorbing sound, thereby enhancing the acoustic performance of ceramic-tiled walls by reducing noise and minimizing sound reflection, all while preserving their aesthetic appearance.

To optimize the acoustic performance, we performed simulations considering all structural parameters, with the constraint of maintaining thin cavities and added layers to ensure practical integration into existing walls. To this end, we also incorporated wooden beams with cross-sectional dimensions of $1~\text{cm} \times 1~\text{cm}$, which serve both as connectors and as supports to bond the ceramic sub-tiles to the wall.
\begin{table}[h!]
    \centering
    \caption{Material parameters}
    \begin{tabular}{c c c c c c}\hline
         &E (GPa)&$\rho$ ($\mathrm{kg/m^3}$)&$\nu$&$\mu$ (Pa.s)&v ($\mathrm{m/s}$)\\\hline
         Ceramic& 55&3000&0.3&-&4967 \\
         Air&-&1.2&-&$1.86\times 10^{-5}$&343\\\hline
    \end{tabular}
    \label{tab1}
\end{table}
\subsection{Numerical model}
We conduct numerical simulations using the commercial finite element software COMSOL Multiphysics, utilizing the Thermoviscous Acoustics Module. Using this module, we solve the full set of linearized governing equations for compressible, viscous, and thermally conducting fluids in the frequency domain. These include the linearized Navier–Stokes equation \ref{eq1} for momentum conservation, the continuity equation \ref{eq2} for mass conservation, and the energy equation \ref{eq3} for thermal effects:
\begin{equation}\label{eq1}
-i \omega \rho_0 \mathbf{v} = -\nabla p + \eta \nabla^2 \mathbf{v} + \left( \zeta + \frac{1}{3} \eta \right) \nabla (\nabla \cdot \mathbf{v}),
\end{equation}  
\begin{equation}\label{eq2}
-i \omega \rho = -\rho_0 \nabla \cdot \mathbf{v},
\end{equation}
\begin{equation}\label{eq3}
-i \omega \rho_0 C_p T + i \omega \alpha_p T_0 p = \nabla \cdot (k \nabla T),
\end{equation}
where $\nabla$ denotes the gradient operator, $\mathbf{v}$ is the acoustic velocity vector, $p$ is the acoustic pressure, $\rho$ is the density, $T$ is the temperature, and $\omega$ is the angular frequency. The parameters $\rho_0$, $T_0$, $\eta$, $\zeta$, $C_p$, $\alpha_p$, and $k$ denote the equilibrium density, ambient temperature, shear viscosity, bulk viscosity, specific heat at constant pressure, thermal expansion coefficient, and thermal conductivity, respectively.

We point out that there is no transmission through the air–ceramic interface, which is modeled as a hard boundary condition ($-\textbf{n} \cdot \frac{1}{\rho_0} \nabla {p} = 0$), where $\textbf{n}$ is the outward unit vector to the surface. Other interfaces are subject to quasi-periodic (Floquet-Bloch) and wall (hard boundary) conditions.

The finite element model is schematically shown in \autoref{Figure1}(b). To simulate an incident wave, we used a Background Pressure Field applied to the air region in front of the ceramic wall structure and in the air cavity. A normally incident plane wave with unit amplitude is applied along the negative $x$-direction. 

The tuning cavity structures are fabricated by ceramic material which has a significantly higher acoustic impedance than air. Due to the large impedance mismatch and negligible structural vibration, the cavity walls are defined as rigid, and hard-wall boundary conditions are applied around the cavity. If the walls are compliant, coupling effects should be taken into account in the acoustic–solid interaction module. Besides, periodic boundary conditions are applied to the lateral sides of the air domain. To resolve the viscous boundary layers, the maximum mesh size inside the slits is set $1/6$ to the minimum mesh size. For the air cavities, the maximum element size is kept below $1/8$ of the shortest wavelength in the target frequency range. In addition, the scattered pressure is measured on a surface just in front of the ceramic tuning cavity. The reflection coefficient $r$ is obtained as the ratio between the scattered and incident acoustic intensity. Then, the absorption coefficient is finally computed using the complex reflection coefficient $r$, extracted from the average sound pressure at the incident boundary as $\alpha = 1 - |r|^2$.

\subsection{Analytical model}
In this section, we derive the equivalent model of a cavity tuning as an electrical circuit, and calculate its absorption coefficient for a single resonator and for a series of resonators with different cavity's volumes, respectively. Our main methodology is to correlate mechanical acoustic parameters to the electrical model and incorporate thermoviscous losses. The sound wave is assumed to be harmonic, with time and frequency dependence given by $ p= A e^{j \omega t} $. In general, the tuning cavity can be modeled as a lumped RLC circuit model. The equivalent electrical circuit of the system is shown in \autoref{Figure1}(c), with the analogy: inductance $ L $ represents the acoustic mass of the air in the slit; capacitance $ C $ represents the compliance of the cavity volume; resistance $ R $ accounts for Thermoviscous losses in the slit and cavity. 

The impedance of a tuning cavity is given by
$
Z_{\rm cavity} = R + j\omega L +\frac{1}{j\omega C}
$, where $L= \dfrac{\rho_0 d}{S}$ is the effective inductance, $ C = \dfrac{V}{\rho_0 v^2} $ is the compliance. Here $ \omega $ is the angular frequency and $j^2=-1$, $ \rho_0 $ is the air density, $d$ is the slit depth, $ S=e\; l $ is the cross-sectional area of the slit, $ v $ is the speed of sound, and $ V=b\; c \; l $ is the cavity volume with $l$ denoting the length of the subtiles, as depicted in \autoref{Figure1}(a).

Thermoviscous losses are included in the real part $ R $ of the impedance. These losses are due to viscous friction of air particles near the walls of the neck. An empirical expression of $ R $ considering these effects is given as
$
R = \dfrac{8 \mu d}{\pi S^2},
$
where $ \mu $ is the dynamic viscosity of air.

The absorption coefficient is defined by the impedance mismatch between the resonator and the surrounding medium:
\begin{equation}
   \alpha = 1 - \left| \frac{Z_0 - Z_{\rm cavity}}{Z_0+ Z_{\rm cavity}} \right|^2=\frac{4 R}{\left| Z_{0} + Z_{\rm cavity} \right|^2}, 
\end{equation}
where $ Z_0 = \rho_0 c $ is the characteristic impedance of air.

Now, we consider $ N $ tuning cavities placed in parallel with varying cavity's volumes $ V_i=e\, c\, b_i$. For each resonator we have
$
L = \frac{\rho_0 \, d}{S}$ and $\
C_i = \frac{V_i}{\rho_0 v^2}
$, and the absorption coefficient for the $ i $-th resonator is:
$$
\alpha_i = \frac{4 R}{\left| Z_{0} + Z^{i}_{\rm cavity} \right|^2}
,$$
where $
Z^{i}_{\rm cavity}$ is the impedance of the $ i $-th resonator. The total absorption coefficient is obtained by the sum:
$
\alpha_{\text{total}} = \sum_{i=1}^{N} \alpha_i
$.
\subsection{Samples and experiments}
\begin{figure}[h!]
    \centering
    \includegraphics[width=0.5\linewidth]{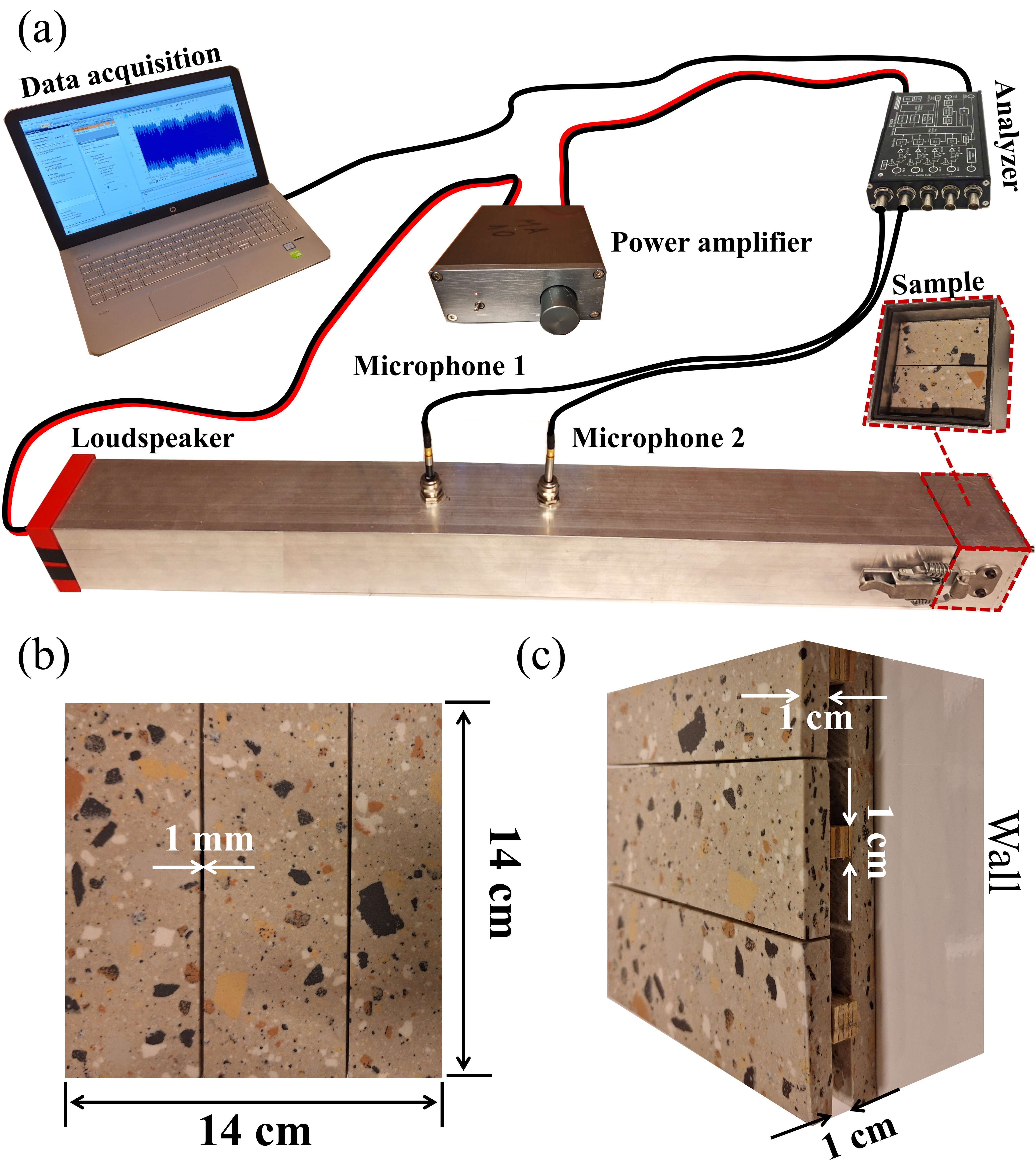}
    \caption{Illustration of the experimental system. (a) The experimental setup. The noise signal is generated by a computer that runs a home-made MATLAB code, and is transferred sequentially to the signal analyzer, the amplifier and the loudspeaker. The generated acoustic pressure is measured by two microphones, both of which are connected to the signal analyzer. The recorded signals are then processed using MATLAB, where a Fast Fourier Transform is applied to analyze the sound absorption coefficient. Photography of the ceramic metatile wall: (b) front view and (c) side view.}
    \label{Figure4}
\end{figure}

The experimental setup is based on a custom-built impedance tube system that employs the two-microphone Transfer Function method prescribed in the ISO 10534-2 standard, as depicted in \autoref{Figure4}(a). This system is utilized to determine both the complex reflection coefficient $r$ and the dimensionless sound absorption coefficient $\alpha$ of the tuning cavity under normal incidence at various frequencies \cite{baali2023design}. 

Ceramic tiles were used as the base material for the subtiles which were then connected to the wall by wooden beams. The ceramic tiles, supplied by IRIS Company, were experimentally characterized through static and dynamic tests under small strain conditions. The measured elastic properties are as follows: Young’s modulus $E_c=(55\pm5)$\,GPa, Poisson’s ratio $\nu_c=0.3$, and density $\rho_c=(3000\pm300)$\,kg/m$^3$ \cite{lemkalli2025controlling}. The complete specimen is shown in \autoref{Figure4}(b) and (c), with overall dimensions of $14\times14\times1$\,cm$^3$.

The apparatus consists of a rigid, rectangular impedance tubes with internal cross-sectional area of $7~\text{cm} \times 7~\text{cm}$ and $14~\text{cm} \times 14~\text{cm}$ and a total length of $60~\text{cm}$ and $120~\text{cm}$, respectively. A broadband loudspeaker is mounted at one end of the tubes, while the ceramic metatile sample is installed at the opposite end. Two identical microphones are positioned along the side of the tubes at distances of $x_1=20~\text{cm}$ and $x_2=30~\text{cm}$ from the surface of the tuning cavity structures. These microphones will capture the acoustic pressure corresponding to the incident and reflected waves within the tube. 
The loudspeaker emits a broadband random signal spanning the frequency range of $50$ \si{Hz} to $1500$ \si{Hz}. The acoustic pressure at each microphone position is used to solve the pressure field in the tube, which can be modeled as:
$p(x) = A e^{-j k x} + B e^{j k x}$,
where $A$ and $B$ denote the amplitudes of the incident and reflected waves, respectively, $k$ is the acoustic wavenumber, and $x=0$ corresponds to the position of the ceramic metatile cavity. The microphone signals are amplified and digitized using a high-precision, multi-channel dynamic signal acquisition module connected to the computer for further data processing and analysis, see \autoref{Figure4}(a).

By computing the transfer function $H_{12}= \frac{p(x_2)}{p(x_1)}$ between the two microphone signals, one can solve for the complex reflection coefficient 
\begin{equation}
  r=\frac{{H}_{12}{e}^{{jk}{x}_{1}}-{e}^{{jk}{x}_{2}}}{{e}^{-{jk}{x}_{2}}-{H}_{12}{e}^{{jk}{x}_{1}}},   
\end{equation}
and subsequently deduce the sound absorption coefficient using the relation:$ \alpha = 1 - |r|^2.$

\section{Results and discussion}
\subsection{Acoustic response of the standard ceramic tile}
\begin{figure}[h!]
    \centering
    \includegraphics[width=0.6\linewidth]{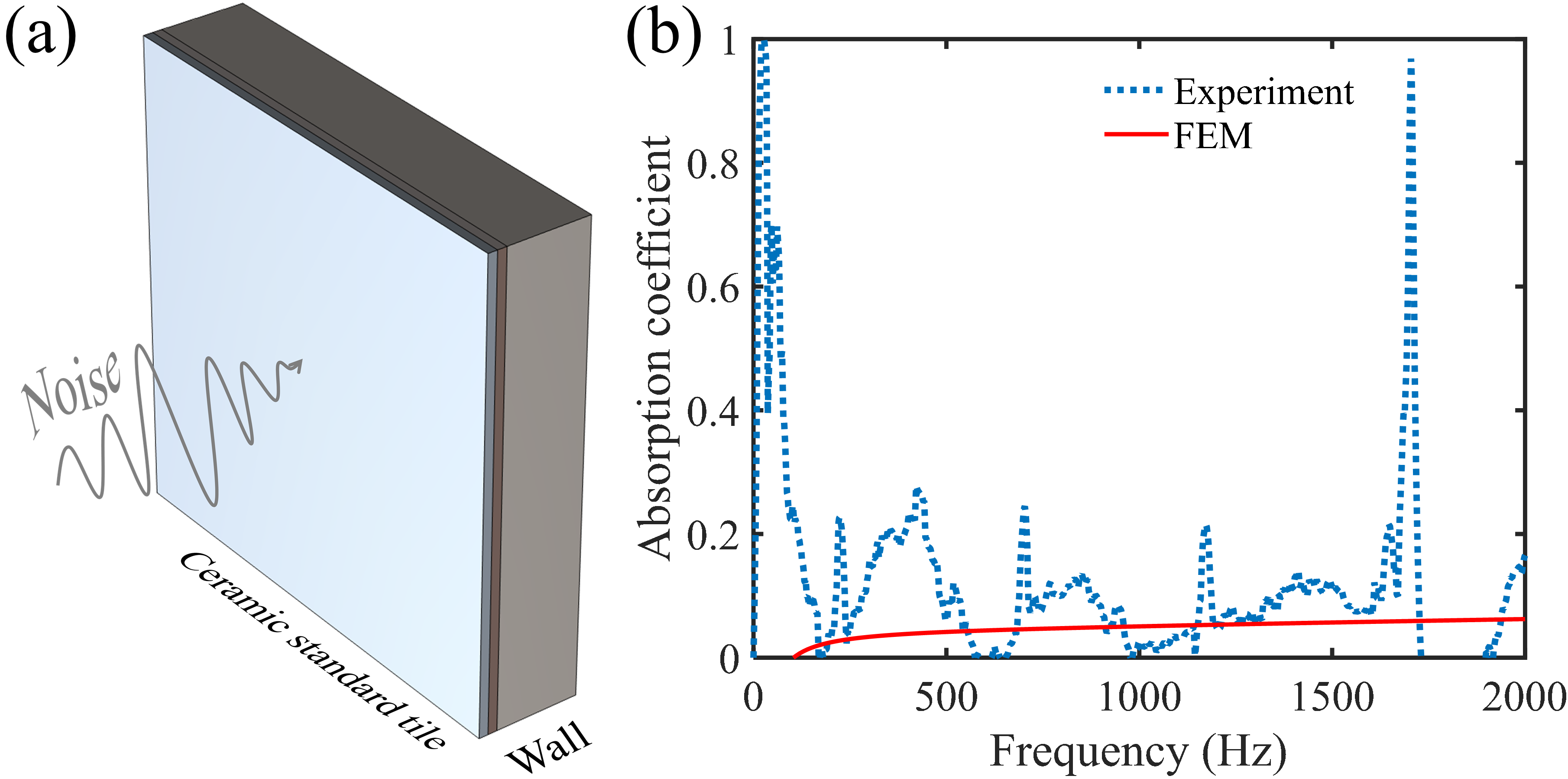}
    \caption{Sound absorption coefficient of the standard ceramic tile. (a) Schematic illustration of the ceramic tile. (b) Numerical and experimental responses of the sound absorption coefficient as a function of frequency.%\textcolor{blue}{Please explain striking discrepancy between experiment and FEM} 
    }
    \label{figure7}
\end{figure}

To better understand the effect of integrating tuned cavity structures into tiled walls, we begin by experimentally and numerically investigating the sound absorption coefficient of a standard ceramic tile, as shown in \autoref{figure7}(a). It is well known that standard ceramic tiles exhibit high sound wave reflectivity, in other words, they reflect much of the noise generated by the surrounding environment. Furthermore, we present in \autoref{figure7}(b) the sound absorption coefficient as a function of frequency for the standard ceramic tile. It is clear that, in their conventional form, ceramic tiles provide negligible sound absorption and are predominantly reflective. This result serves as a reference for comparison with the subsequent sections, where we transform the standard tiles into metatiles by incorporating tuned cavity resonators based on the Helmholtz resonator. \textcolor{black}{There is a noticeable discrepancy between the experimental and numerical results. In the numerical simulations, we model an infinite ceramic tile by applying periodic boundary conditions, which idealizes the response of a continuous tiled surface. However, in the experimental setup, we use a small tile sample placed inside an impedance tube. Due to the limited sample size, unavoidable gaps exist between the sample edges and the tube walls. These gaps can behave like quarter-wave resonators, introducing additional resonance peaks in the measured absorption spectrum. Furthermore, the dimensions of the impedance tube define the operational frequency range for sound absorption measurements, which is approximately from  $200$ \si{Hz} to $1600$ \si{Hz}. Outside this range, especially beyond the upper and lower frequency limits, the absorption results become unreliable. This explains the discrepancy between the numerical and experimental results, particularly the additional peaks observed in the experimental data.}

\subsection{Parametric optimization of the tuning cavity}
\begin{figure*}[h!]
    \centering
    \includegraphics[width=1\linewidth]{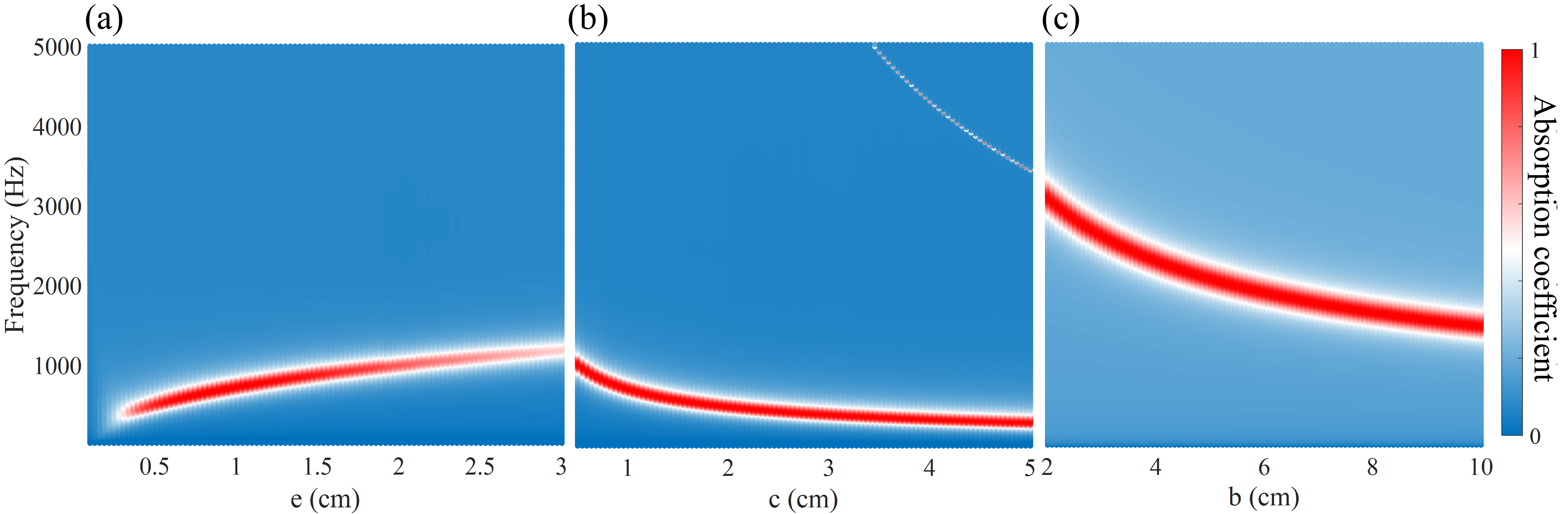}
    \caption{Absorption coefficient as a function of three key cavity parameters. (a) Thickness of the slit e, varied within the range 0.1 to 3 \si{cm}. (b) Width of the cavity c, varied within the range 0.5 to 5 \si{cm}. (c) Length of the cavity b, varied within the range 2 to 10 \si{cm}.}
    \label{Figure2}
\end{figure*}
Considering the requirement of thinner structures in practical building applications, we optimize the geometrical parameters of the tuning cavity, including the slit width $e$, the cavity width $c$, and the cavity length $b$. Additionally, we employ the numerical model illustrated in \autoref{Figure1}(b), using subtiles of length $l = 30$ \si{cm} along the $y$-direction.

We begin by varying the slit width $e$ from $0.1$ mm to $3$ mm, while keeping $c = 1$ \si{cm} and $b = 4$ \si{cm}. As shown in \autoref{Figure2}(a), incorporating a slit and a cavity enables efficient absorption of sound waves over the frequency range from $310$ \si{Hz} to $1290$ \si{Hz}, depending on the variation of $e$. Furthermore, as the slit width increases (i.e., the neck becomes thicker), the absorption peak shifts toward higher frequencies, and the overall absorption coefficient decreases. This behavior can be explained by the Helmholtz resonance mechanisms, which describe the phenomenon of air resistance in the neck of the cavity. With the increment of the neck width of the Helmholtz resonator, acoustic waves can propagate through the neck with less resistance, resulting in a shift of the resonance frequency to higher values and a reduction in the overall sound absorption.

Then we fix the slit width ($e = 1$ \si{cm}) and the cavity length ($b = 4$ \si{cm}), and vary the cavity width $c$ from $0.5$ \si{cm} to $5$ \si{cm}, as shown in \autoref{Figure2}(b). As the cavity width increases, the absorption peak shifts toward lower frequencies, reaching as low as 250 \si{Hz} when $c = 5$ \si{cm}, where nearly total absorption is observed. This indicates that increasing the cavity volume can enhance low-frequency absorption, but meanwhile results in larger thickness which is not aligned with our primary objective of maintaining compact and thin.

Finally, we study the effect of cavity length $b$ by fixing the slit width ($e = 1$ \si{mm}) and the cavity width ($c = 1$ \si{cm}). This group of parameter keeps the structure thin and small enough to be adaptable for tiled wall applications. We vary $b$ from $2$ \si{cm} to $10$ \si{cm} to optimize the design by adjusting only this parameter. As illustrated in \autoref{Figure2}(c), the absorption coefficient shifts toward lower frequencies as the cavity length increases. This indicates that we can optimize our design and maintain a thin profile by tuning the cavity length to achieve broadband absorption at low frequencies, while also reducing the reflectivity of ceramic tiled walls.

\subsection{Oblique Incident Absorption Performance}

\begin{figure*}[h]
    \centering
    \includegraphics[width=0.8\linewidth]{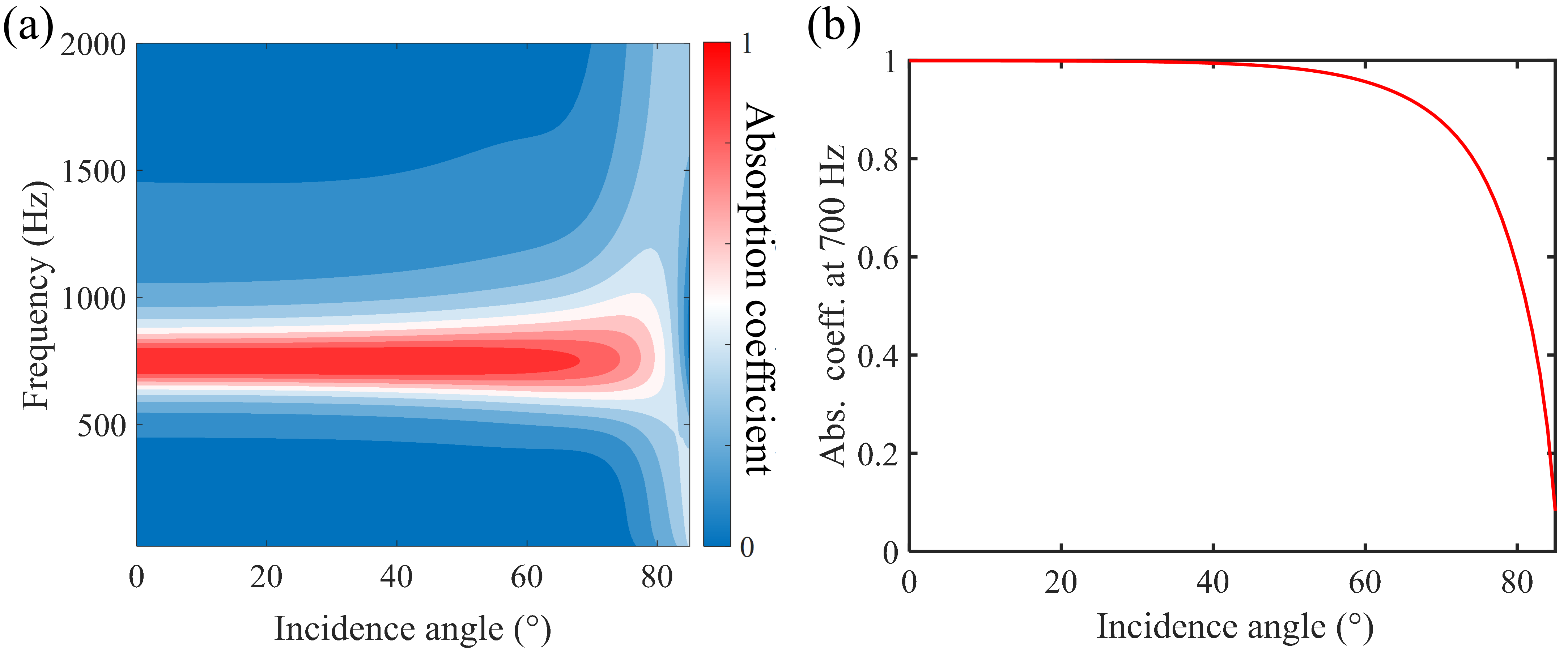}
    \caption{(a) Absorption performance of oblique incident waves. (b) Absorption coefficient at 700 \si{Hz} as a function of the incidence angle. }
    \label{Figure3}
\end{figure*}

We fix an optimal set of geometry parameters as $e = 1$ \si{mm}, $c = 1$ \si{cm} and $b = 4$ \si{cm}, and study the effect of the incident wave angle on the sound absorption coefficient. The results in \autoref{Figure3}(a) demonstrate that the optimized ceramic metatile wall achieves consistently high sound absorption over a wide range of incidence angles. In particular, it maintains absorption values exceeding $0.9$ for incidence angles between $0^\circ \leqslant \theta_i \leqslant 75^\circ$, clearly outperforming conventional acoustic absorbers, as depicted in \autoref{Figure3}(b).

Locally resonator absorbers are characterized by surface impedance that are independent of the incident angle. This simplified behavior can effectively model various types of absorbers, including quarter-wavelength and Helmholtz resonators (tuning cavities), as implemented in our work. However, when such materials are optimized for peak performance at grazing incidence, their absorption efficiency tends to decline at lower angles due to the strong angular dependence of the radiation impedance.

\subsection{Experimental Verifications}

\begin{figure}[hp!]
    \centering
    \includegraphics[width=0.5\linewidth]{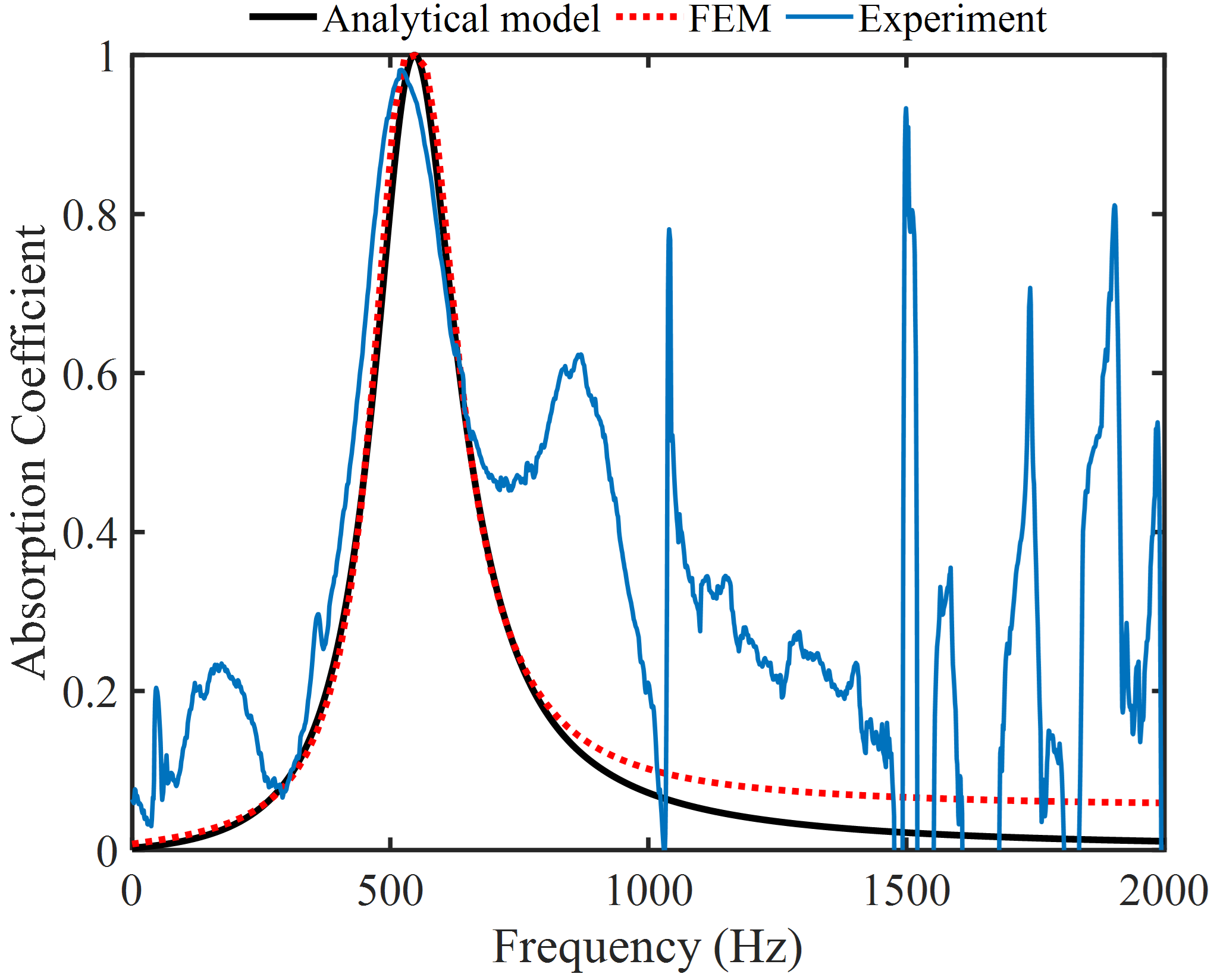}
    \caption{Absorption coefficients of the metatile obtained through experimental measurements, FEM numerical simulations, and analytical modeling.}
    \label{Figure5}
\end{figure}

The results from sound absorption testing are shown in \autoref{Figure5}, where we observe a high degree of consistency with the acoustic absorption coefficient predicted by FEM and the analytical model. This close agreement confirms the reliability of the FEM and analytical approaches in accurately capturing the acoustic absorptive behavior of the ceramic metatile cavity structure. The band sound absorbing structure demonstrates excellent performance within the frequency range of 480 \si{Hz} to 700 \si{Hz}. Remarkably, even with a single tuning cavity defined by the geometric parameters $e = 1~\mathrm{mm}$, $c = 1$ \si{cm}, and $b = 5$ \si{cm}, the structure achieves strong sound absorption in this frequency range, with an average absorption coefficient exceeding $0.7$ and reaching up to 1, thereby achieving quasi-perfect sound wave absorption. The minor discrepancies observed between the experimental and numerical results are attributed to fabrication imperfections, such as dimensional tolerances and surface roughness. Additionally, peaks appearing beyond $1000$ \si{Hz} are likely due to diffraction phenomena caused by the finite size of the sample. It is worth noting that the tested sample measured $6.9~\mathrm{cm} \times 6.9 ~\mathrm{cm}$, closely matching the $7~\mathrm{cm} \times 7~\mathrm{cm}$ cross-section of the impedance tube used in the measurements.

\subsection{Coupling of multiple cavities}
\begin{figure}[h]
    \centering
    \includegraphics[width=0.5\linewidth]{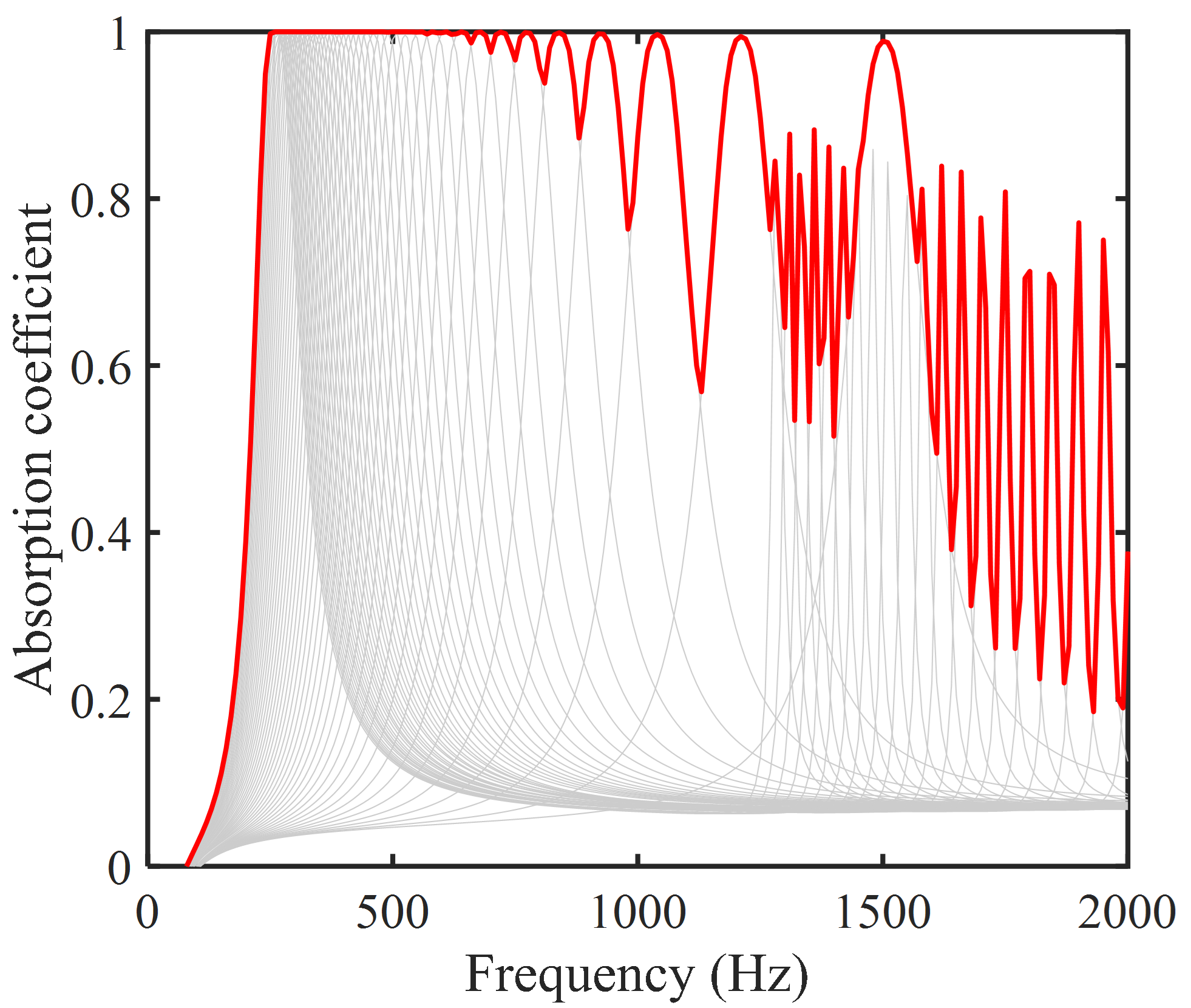}
    \caption{Absorption coefficients of the ceramic metatile cavity obtained from FEM numerical simulations for multiple tuning cavities with varying cavity lengths $b$ ranging from $1~\mathrm{cm}$ to $28~\mathrm{cm}$.}
    \label{Figure 7}
\end{figure}

After validating our numerical model with experimental results, we proceed to study the effect of coupling multiple cavities to open a broadband acoustic band gap at low frequencies. It is important to recall that our ceramic metatile design primarily targets thin wall applications. To preserve the aesthetic appearance of ceramic-tiled walls, we keep the slit width $c$ fixed at $1~\mathrm{cm}$, along with the neck dimensions. To explore the effect of cavity depth, we vary the parameter $b$ from $1~\mathrm{cm}$ up to $28~\mathrm{cm}$, considering that standard ceramic tiles can reach up to $30~\mathrm{cm}$ in thickness, thus enabling the formation of deep cavities behind the tiles. The findings of this study are presented in \autoref{Figure 7}. The results show that coupling large cavities of 28 \si{cm} with smaller ones as shallow as $1~\mathrm{cm}$ can effectively generate broadband sound absorption within the frequency range of $200$ \si{Hz} to $1000$ \si{Hz}. This performance is notably superior compared to existing designs in the literature, achieving excellent absorption while maintaining a thin profile and preserving the visual appeal of ceramic-tiled walls.

\begin{figure*}[ht!]
    \centering
    \includegraphics[width=1\linewidth]{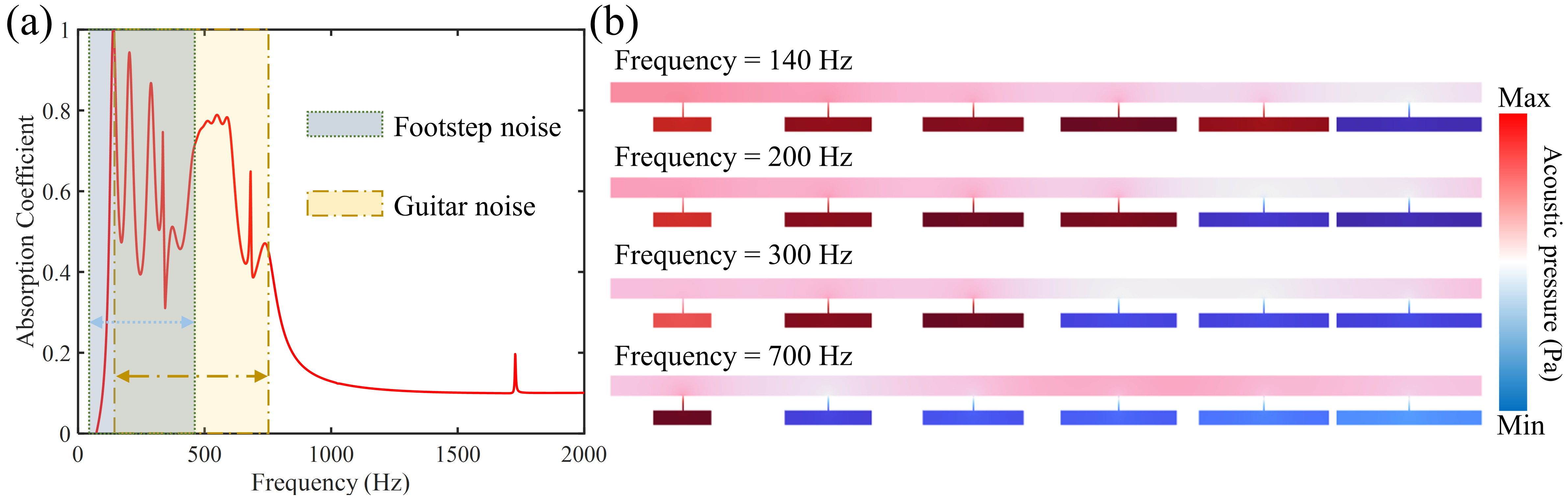}
    \caption{Rainbow trapping sound absorption in the supercell composed of ceramic metatile cavities. (a) Absorption coefficient as a function of frequency, with colored zones indicating the most common indoor noise frequency ranges. (b) Acoustic pressure field screenshots of the cross-section $xz$-plane within the supercell structure. The parameter b takes values of $14$ \si{cm} (425 \si{Hz}), $16$ \si{cm} (398 \si{Hz}), $17$ \si{cm} (386 \si{Hz}), $18$ \si{cm} (375 \si{Hz}), $19$ \si{cm} (365), and $20$ \si{cm} ( 350 \si{Hz}), with l fixed at $30$ \si{cm}.}
    \label{Figure 8}
\end{figure*}

Building on the understanding that varying the cavity length can enhance sound absorption while maintaining a thin design using reflective ceramic tiles, we design a superlattice composed of multiple tuning cavities with different lengths. This super-lattice consists of $6$ unit cells, each with the same slit width $e = 1~\mathrm{mm}$ and cavity width $c = 1~\mathrm{cm}$, while the cavity length $b$ varies across the cells, the geometrical parameters are based on a previous study and are selected to achieve low-frequency broadband sound absorption. The parameter b takes values of $14$ \si{cm}, $16$ \si{cm}, $17$ \si{cm}, $18$ \si{cm}, $19$ \si{cm}, and $20$ \si{cm}, with l fixed at $30$ \si{cm}. The corresponding frequencies, calculated using the expression: $\frac{v}{2 \pi}\sqrt{\frac{e\, l}{c\, b \, l \, d}}$ 
are approximately 425 \si{Hz}, 398 \si{Hz}, 386 \si{Hz}, 375 \si{Hz}, 365 \si{Hz}, and 350 \si{Hz}, respectively. The goal is to achieve broadband sound absorption at low frequencies and demonstrate that modifying the cavity volumes within ceramic metatile walls significantly affects the absorption range, making these walls more acoustically absorptive despite being made of reflective ceramic materials. This concept is related to the well-known rainbow trapping phenomenon in acoustic metamaterials \cite{jimenez2017rainbow}. In our study, we refer this mechanism to show that it is possible to achieve low-frequency absorption simply by varying cavity lengths while preserving the thin form factor and aesthetic of ceramic-tiled walls, as presented in \autoref{Figure 8}.

By increasing the parameter $b$, broadband sound absorption can be achieved in the frequency range of approximately $123$ \si{Hz} to $744$ \si{Hz}, as shown in \autoref{Figure 8}(a). When $6$ thin cavities are assembled into a single superlattice, the largest cavity produces a strong resonance peak around $140$ \si{Hz}, where the absorption coefficient reaches a value of $1$. Following this peak, the absorption decreases slightly as additional cavities begin to activate, as illustrated in \autoref{Figure 8}(b). The absorption remains effective up to around $740$ \si{Hz}, after which it returns to a more typical level once all cavities have been excited. These results indicate that arranging ceramic metatiles to create a linear variation in cavity volume effectively reduces sound reflection by achieving absorption coefficients of $0.7$ or higher over a broad frequency band ranging from $140$ \si{Hz} to $744$ \si{Hz}.

To summarize, we have demonstrated the effectiveness of incorporating tuning cavities into ceramic tile walls to enhance their acoustic performance. 

This approach significantly enhances the tiles' ability to absorb noise commonly generated in indoor environments, such as footsteps (typically $30$ \si{Hz} to $470$ \si{Hz}), musician instruments (like guitar within $140$ \si{Hz} to $740$ \si{Hz}) and sounds from household appliances typically occurring within the $80$ \si{Hz} to $1000$ \si{Hz} frequency range \cite{li1991perception,serafin2009extraction}, which are usually reflected by conventional tiled surfaces. Our findings show that low-frequency noise can be efficiently absorbed by simply adjusting the volume of the cavity. While our study focused on a configuration with six cavities, real-world applications could involve a larger number of cavities with lengths varying from $1~\mathrm{cm}$ to $28~\mathrm{cm}$, as illustrated in \autoref{Figure 7}. Moreover, we emphasize that our design maintains a slim profile, with each cavity having a width of only $1~\mathrm{cm}$, an important consideration for practical use in building applications where space efficiency is critical.

\section{Conclusion}
To conclude, we analytically, numerically, and experimentally demonstrated the effectiveness of the proposed physical mechanism and its significantly added value compared to previous studies on ceramic tiled walls. First, we highlighted the acoustic performance of ceramic metatile walls by investigating the effect of cavity tuning. In particular, we introduced an acoustic metamaterial based on a Helmholtz resonator capable of achieving near-perfect absorption when integrated into a ceramic tiled wall. We showed that increasing the cavity length shifts the absorption peaks to lower frequencies, while increasing the neck thickness of the Helmholtz resonator shifts the peaks to higher frequencies and reduces the absorption coefficient. Furthermore, the structure maintains its sound absorption behavior even under oblique incidence of acoustic waves. Finally, we demonstrated the effectiveness of rainbow trapping in ceramic metatile cavities, enabling broadband and low-frequency sound absorption. The findings of this study open a new avenue for research on ceramic-tiled surfaces, particularly tiled walls which typically suffer from high sound reflectivity.

\section*{Acknowledgments}
This research was funded by IRIS, BTU Trust, MetaMAT Ltd UK, the French ANR (contract "ANR-21-CE33-0015") and the European Union under Marie Skłodowska-Curie Actions Postdoctoral Fellowships (No. 101149710). The authors acknowledge the support of the French-Swiss SMYLE Network. 
We thank Julien Joly (Université Marie et Louis Pasteur, département d'Electronique) for the 3D printed structures.

\bibliography{biblio}

\end{document}